\documentclass[8.5pt,twoside,twocolumn]{article}
\usepackage[super,sort&compress,comma]{natbib} 
\usepackage{mhchem}
\usepackage{times,mathptmx}
\usepackage{sectsty}
\usepackage{balance} 

\usepackage{graphicx} 
\usepackage{amsmath}    
\usepackage{verbatim}   
\usepackage{color}      
\usepackage{subfigure}  
\usepackage{hyperref}   
\usepackage{lipsum}
\usepackage{lastpage}
\usepackage[format=plain,justification=raggedright,singlelinecheck=false,font=small,labelfont=bf,labelsep=space]{caption} 
\usepackage{fancyhdr}

\begin{document}

\thispagestyle{plain}
\fancypagestyle{plain}{
\fancyhead[L]{\includegraphics[height=8pt]{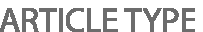}}
\fancyhead[C]{\hspace{-1cm}\includegraphics[height=20pt]{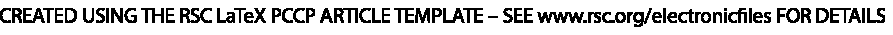}}
\fancyhead[R]{\includegraphics[height=10pt]{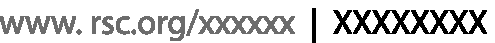}\vspace{-0.2cm}}
\renewcommand{\headrulewidth}{1pt}}
\renewcommand{\thefootnote}{\fnsymbol{footnote}}
\renewcommand\footnoterule{\vspace*{1pt}%
\hrule width 3.4in height 0.4pt \vspace*{5pt}} 
\setcounter{secnumdepth}{5}

\makeatletter 
\def\subsubsection{\@startsection{subsubsection}{3}{10pt}{-1.25ex plus -1ex minus -.1ex}{0ex plus 0ex}{\normalsize\bf}} 
\def\paragraph{\@startsection{paragraph}{4}{10pt}{-1.25ex plus -1ex minus -.1ex}{0ex plus 0ex}{\normalsize\textit}} 
\renewcommand\@biblabel[1]{#1}            
\renewcommand\@makefntext[1]%
{\noindent\makebox[0pt][r]{\@thefnmark\,}#1}
\makeatother 
\renewcommand{\figurename}{\small{Fig.}~}
\sectionfont{\large}
\subsectionfont{\normalsize} 

\fancyfoot{}
\fancyfoot[LO,RE]{\vspace{-7pt}\includegraphics[height=9pt]{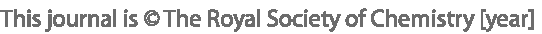}}
\fancyfoot[CO]{\vspace{-7.2pt}\hspace{12.2cm}\includegraphics{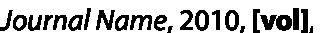}}
\fancyfoot[CE]{\vspace{-7.5pt}\hspace{-13.5cm}\includegraphics{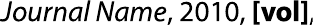}}
\fancyfoot[RO]{\footnotesize{\sffamily{1--\pageref{LastPage} ~\textbar  \hspace{2pt}\thepage}}}
\fancyfoot[LE]{\footnotesize{\sffamily{\thepage~\textbar\hspace{3.45cm} 1--\pageref{LastPage}}}}
\fancyhead{}
\renewcommand{\headrulewidth}{1pt} 
\renewcommand{\footrulewidth}{1pt}
\setlength{\arrayrulewidth}{1pt}
\setlength{\columnsep}{6.5mm}
\setlength\bibsep{1pt}

\twocolumn[
  \begin{@twocolumnfalse}
\noindent\LARGE{\textbf{Quantum interference and contact effects in dangling bond loops on H-Si(100) surfaces}}
\vspace{0.6cm}

\noindent\large{\textbf{Andrii Kleshchonok,$^{\ast}$\textit{$^{a}$} Rafael Guti\'{e}rrez,\textit{$^{a}$},   and
Gianaurelio Cuniberti\textit{$^{a,b}$}   
}}\vspace{0.5cm}

\noindent\textit{\small{\textbf{Received Xth XXXXXXXXXX 20XX, Accepted Xth XXXXXXXXX 20XX\newline
First published on the web Xth XXXXXXXXXX 200X}}}

\noindent \textbf{\small{DOI: 10.1039/b000000x}}
\vspace{0.6cm}

\noindent \normalsize{We perform  electronic structure and quantum transport studies of  dangling bond loops created on H-passivated Si(100) surfaces and connected to carbon nanoribbon leads. We model loops with straight and zigzag topologies as well as with varying lenght with an efficient density-functional based tight-binding electronic structure approach (DFTB)  . Varying the length of the loop or the lead coupling position we induce the drastic change in the transmission due to the electron interference. Depending if the constructive or destructive interference within the loop takes place we can noticeably change transport properties by few orders of magnitude. These results propose a way to engineer the closed electronically driven nanocircuits with high transport properties and exploit the interference effects in order to control them. }
\vspace{0.5cm}
 \end{@twocolumnfalse}
  ]

\section{Introduction}
\footnotetext{\dag~Electronic Supplementary Information (ESI) available: [details of any supplementary information available should be included here]. See DOI: 10.1039/b000000x/}


\footnotetext{\textit{$^{a}$~Institute for Materials Science,Dresden University of Technology; andrii.kleshchonok@nano.tu-dresden.de}}
\footnotetext{\textit{$^{b}$~Institute for Materials Science and Max Bergmann Center of Biomaterials, TU Dresden, 01062 Dresden, Germany. Center for Advancing Electronics Dresden, TU Dresden, 01062 Dresden, Germany.Dresden Center for Computational Materials Science (DCCMS), TU Dresden, 01062 Dresden, Germany }}

During the last decades the  miniaturization of electronic devices has steadily approached the atomic scale.  As a result, 
a variety of new challenges arise, including the building of atomic scale circuit elements and logic elements on the basis of individual molecules as well as gaining control over the inter-connects that ensure communication between the individual nanoscale components\cite{joachim2000electronics,Godlewski,review,wolkow2014silicon,Joachim1,Owen,cuevas,0953-8984-22-13-133001}. 
 Hereby,  scanning tunneling microscopy (STM) has revealed as an invaluable tool to manipulate matter at sub-nm lengths\cite{nature2,nature3}, so that it has opened the possibility to engineer physical properties with atomic-scale precision \cite{review}. 
In particular, a promising candidate for nanoelectronic applications  are dangling bond wires (DBW), which are  formed by selectively removing hydrogen atoms from  Si(100) or Ge(100) passivated surfaces with the 
 help of an STM tip; desorption of the H-atoms leaves   behind an unpaired 
electron (a dangling bond)\cite{review}.  This selective engineering at the atomic scale opens the fascinating possibility  to design planar nanocircuits with complex geometry and tunable conduction properties\cite{Joachim_nano,Boland,nature3, Wolkow}. Additionally, DBW are inherently compatible with standard semiconductor based nanotechnologies, so that interfacing both domains may be a viable alternative in a near future. 
In  real  DB wires, however, the charge transport efficiency may be strongly suppressed by Jahn-Teller distortions and buckling of the Si surface atoms \cite{Lee,Bird,nature1,Joachim5,Joachim6}. Theoretical studies addressing the charge transport through DBW mainly deal with charge transport along dimer rows \cite{Joachim_nano,Joachim5}. However, less attention has been paid to other, quasi 2D topologies, which have however been demonstrated experimentally\cite{Soukiassian2003121,Boland}. Although electronic coupling perpendicular to the DB rows is smaller than along the dimer rows, it is non-zero and charge transport along DB loops may show interesting features like quantum interference effects.   Motivated by this, we address in this investigation the problem of the charge transport properties of  realistic  DB loops on Si(100) surfaces coupled to mesoscopic carbon nanoribbons acting as electrodes. We  focus on intrinsic properties of the dangling-bond system  like the loop topology and size and their 
influence on quantum interference effects, whose signatures show up in the quantum mechanical transmission function of the system. Since the DB loops need to be interfaced with the environment to probe charge transport, we also address the influence of the loop-electrode contact geometry and reveal a very sensitive dependence of charge transport on its local atomic structure.

In our study of dangling-bond loops on H-passivated Si(100), we  use a density functional-based tight binding (DFTB) approach \cite{rauls1999stoichiometric,pecchia2008non} to perform  structural relaxation as well as to compute the electronic structure of the different configurations dealt with.   The DFTB approach allows to treat large structures ($\approx 2000$ atoms) with reasonable CPU time, while still keeping enough accuracy to provide a realistic insight into the properties of the system. From these calculations we obtain the corresponding electronic density of states (DOS), the Fermi energy ($E_F$), and the Hamiltonian and Overlap matrix elements. The latter two are then used as input for a Green's function based approach to compute the quantum mechanical transmission $T(E)$ of the system. 
In brief, the standard Landauer-B\"{u}ttiker formalism \cite{datta1997electronic} is used, where the energy-dependent transmission function is computed as $T(E)=Tr(\Gamma_{\textrm{L}}(E) G^{\textrm{r}}(E) \Gamma_{\textrm{R}}(E) G^{\textrm{a}}(E))$. Here, $G^{\textrm{r(a)}}$ are  retarded (advanced) Green's functions of the system, $\Gamma_{\textrm{L,R}}=i\left( \Sigma^{\textrm{r}}_{\textrm{L,R}} - \Sigma^{\textrm{a}}_{\textrm{L,R}} \right)$ are the electrodes spectral densities, and  $\Sigma^{\textrm{r(a)}}_{\textrm{L(R)}}$ are  retarded (advanced) self-energy functions of the left L  (right R) leads, which encode the electronic structure of the electrodes as well as the electronic coupling between electrode states and, in our case, the DB states. The self-energies are calculated with the well-known iterative Lopez-Sancho procedure \cite{sancho1985highly}. 

\begin{figure}[!ht]
\begin{center}
\includegraphics[width=0.5\textwidth]{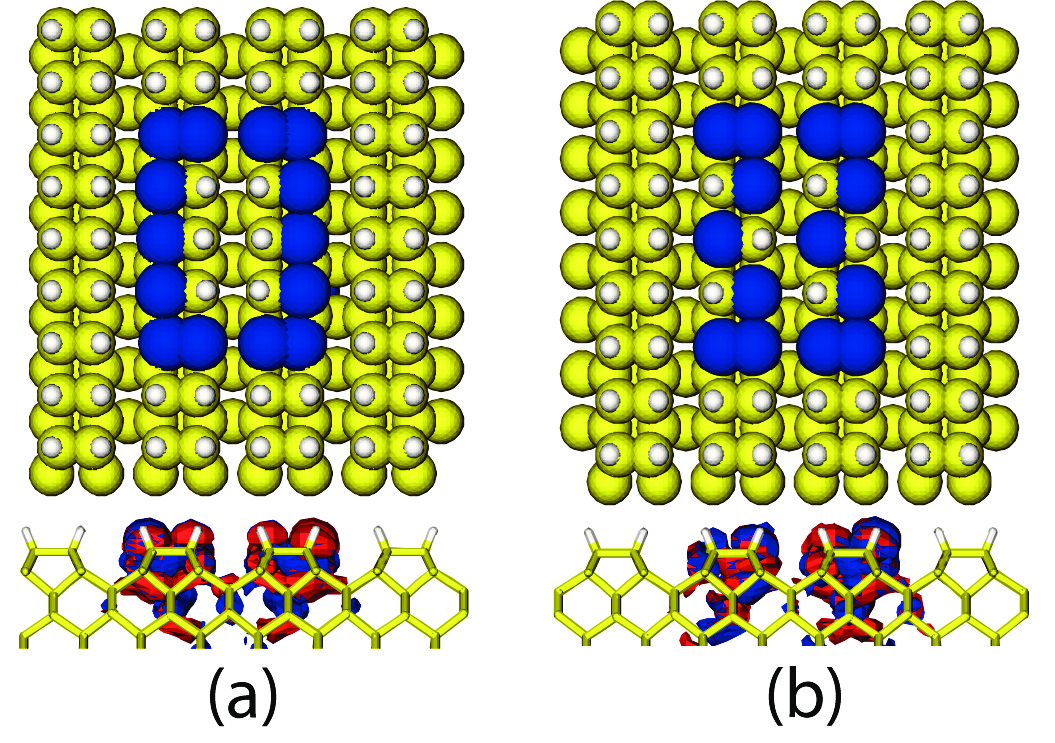}\\
  \caption{ (a) Straight and (b) zigzag DB loops. The upper panels of (a) and (b) show the depassivated atoms highlighted in dark blue  on an otherwise H-passivated Si(100) surface.  The bottom panels of (a) and (b) display side views of the real part of electron wave function of the  { DB states in the Si band gap}, where red color corresponds to the positive and blue - to the negative contributions. Although the largest spectral weight is on the dangling bond related surface states, there is still some contribution from deeper silicon layers.}
  \label{fig_orb}
\end{center}
\end{figure}

\begin{figure*}[hb]
\begin{center}
\includegraphics[width=0.9\textwidth]{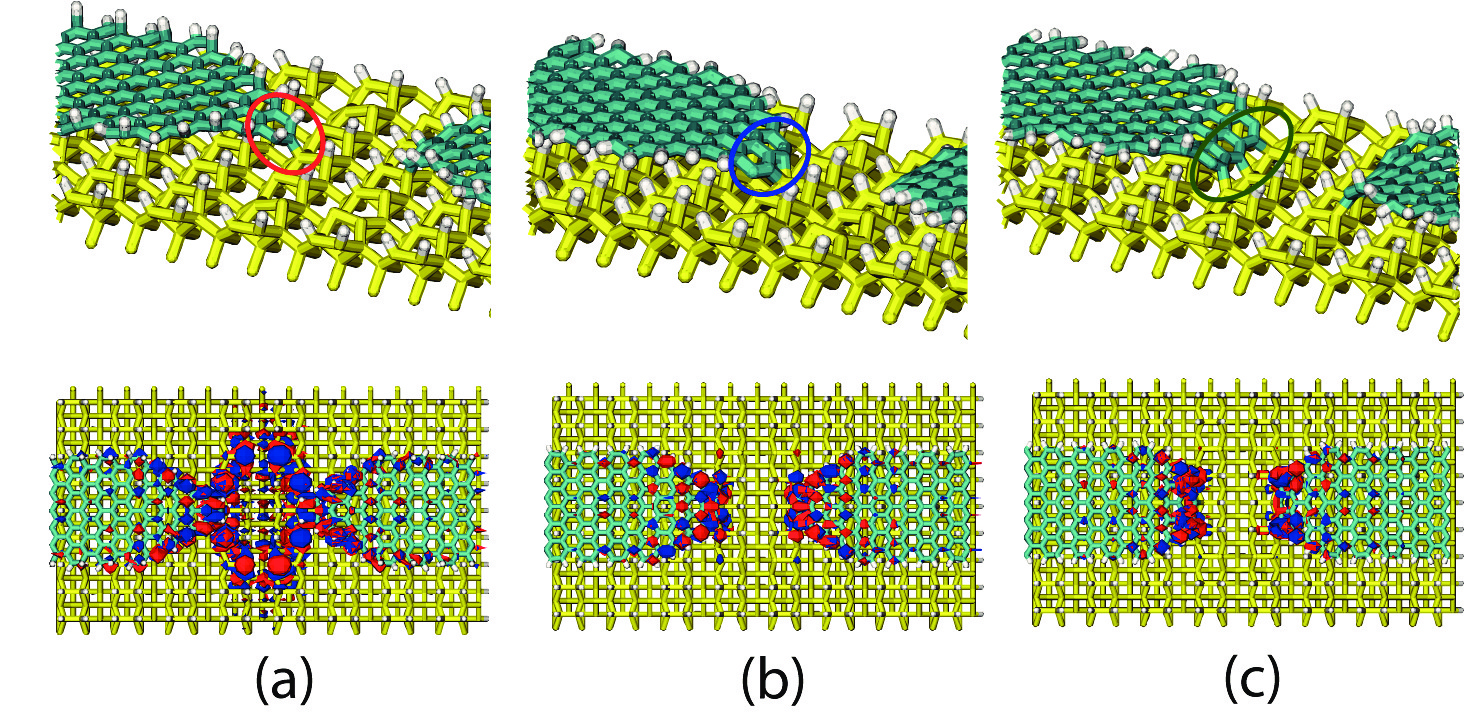}
    \caption{Graphene nanoribbon electrodes terminated with one (a), two (b) and three (c) C atoms and corresponding electronic structure { in an energy window of $\pm 0.2$ eV around the Fermi energy $E_F$}
.}
  \label{fig_syst}
\end{center}
\end{figure*}

\begin{figure}
\begin{center}
\includegraphics[width=0.46\textwidth]{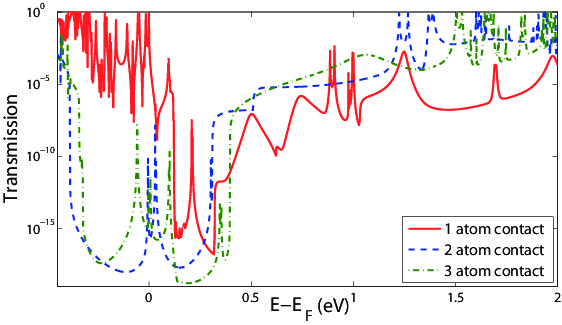}
    \caption{Transmission of a straight DB loop coupled to  graphene nanoribbon electrodes  terminated with (a) one (solid line in red, (b) two (dashed line in blue) and (c) three (dotted-dashed line in green) carbon atoms.}
  \label{fig_trans}
\end{center}
\end{figure}

We focus on two possible topologies of the DB loops, see \ref{fig_orb} for reference: (i) straight, when the Si surface is depassivated along the direction of the dimer row following a straight line, and (ii) zigzag, when  the H atoms are removed in a zigzag way. The last one was predicted to have better charge transport properties in the case of infinite wires \cite{Joachim_nano}. There are two possible orientations of the loop on the surface, but in order to have a higher charge transport efficiency, the longer side of the loop needs to be oriented along the (100) direction, since the electronic coupling perpendicular to the dimer rows is much smaller. During  relaxation the Si DB surface atoms are buckled, showing a Peierls-like distortion, which may additionally reduce the transport efficiency when compared with the ideal unrelaxed loop. In a first step we address loops with a fixed linear dimension,  consisting of five DBs along the dimer rows and four DBs in the perpendicular direction. The charge 
density associated to 
the DB states is localized very close to the Si surface as shown in the lower panel of  \ref{fig_orb}.  
To probe the transport properties of the DB loops, atomic-scale electrical contacts need to be engineered. One possibility is to use metallic gold pads \cite{0953-8984-24-9-095011,0953-8984-23-12-125303}; we have however chosen graphene nanoribbons as shown in \ref{fig_syst}, due to their intrinsic one-atom thickness and the variety of electronic features they display, which may add additional ways to tune electronic transport in hybrid carbon-silicon nanoscale devices.  There exist already experimental studies with molecular nanowires and graphene nanoribbons on  Si surfaces \cite{Kittelmann,Nature_jin,lopinski,xu2011inducing}, so that our proposed transport setup possesses potential experimental relevance.  Due to their planar geometry and one-atom thickness, carbon nanoribbons can directly couple to single Si atoms on the DB loop, while remaining electronically decoupled from the (passivated) Si substrate on which they are lying \cite{xu2011inducing,ritter2009influence,
albrecht2007preferential}. Apart from single atom contacts, it is also possible to  have  trapezoidal terminations of the graphene nanoribbon, i.e. more than one C atom can have a contact to Si atoms on the DB loop. This is also illustrated in panels b) and c) of \ref{fig_syst}).
Naively, one may expect that increasing the number of atoms in the corner of the nanoribbon would improve the electronic coupling to the DB loop and thus increase the conductance (at least at low energies). In fact, the local DOS is found to increase in the contact region (see SI), but at the same time  the Fermi energy of the system is shifted and the spectral weight of the  DB states in the Si band gap is modified, having a mixture of  states coming from DBs and the leads. It turns then out that within a low-energy window $\pm 0.2$ eV around the Fermi level the leads ending with one C atom (\ref{fig_syst}a)  provide a more efficient charge transport pathway than the two other cases with multiple contacts shown in  \ref{fig_syst}b and \ref{fig_syst}c.
This behavior can be clearly seen in \ref{fig_trans}, where we plot  the transmission functions for  the different contact geometries  described above and considered  a straight loop consisting of  five DBs along the dimer rows and four in the perpendicular direction. In the case (a)  we obtain a high and narrow transmission peak near the Fermi energy close to the conduction band, while in the cases (b) and (c) quantum transport is strongly suppressed around the Fermi energy. {The remaining peaks are related to the direct tunnelling between the DB rows and are discussed in SI.}

For the sake of 
clarity, we  limit our following discussion to the case where a single carbon atom of the graphene nanoribbon couples to a single atom on the Si DB loop, which corresponds to  \ref{fig_syst}a. 
The nanoribbon slices for the Lopez-Sancho procedure were taken at a distance of few dimer rows away from the central part $-$including not only Si atoms, but also the first few atoms of the nanoribbon tip$-$, in order to minimize the influence on the DB states. This, together with  periodic boundary conditions, forces us to take quite a large substrate 8 Si atoms thick, 9 atoms width and 8 dimer rows long in the transport direction.

\begin{figure*}[ht]
\begin{center}
\includegraphics[width=0.9\textwidth]{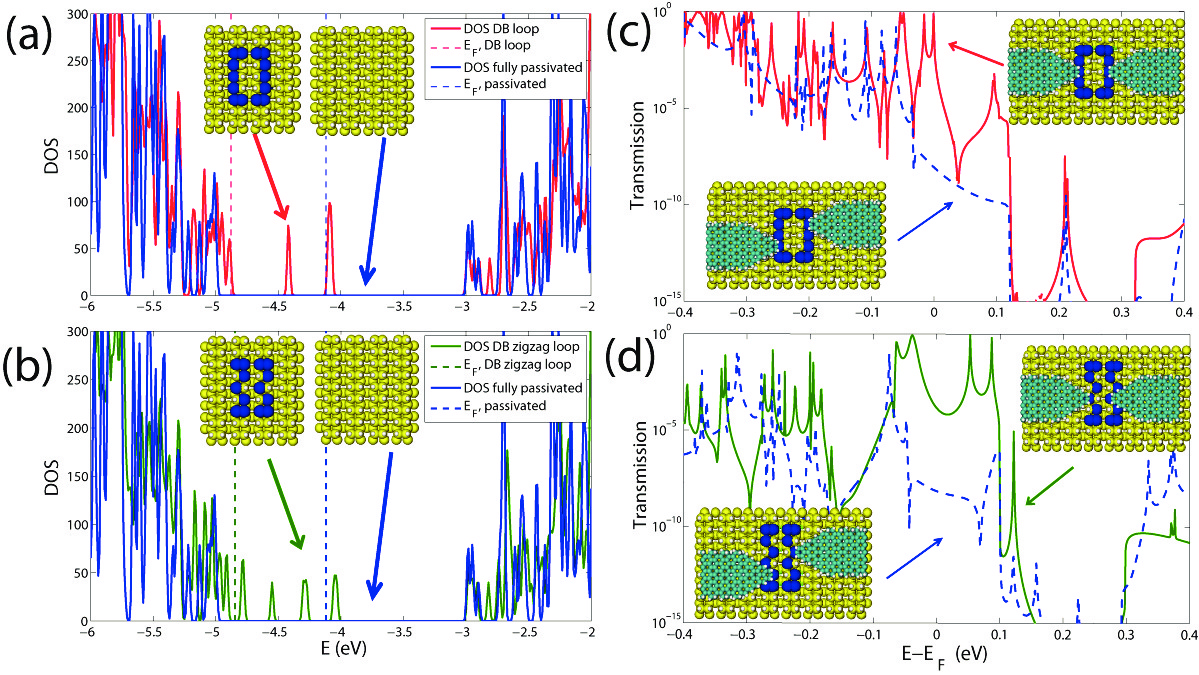}
    \caption{ Left panel: Electronic density of states (DOS) of (a) straight (red curve)  and (b) zigzag (green curve) DB loops. Also shown as reference  the corresponding DOS for the fully H-passivated Si(100) surface. Vertical dotted lines indicate the positions the Fermi level $E_F$ in the different cases.  Right panel: The corresponding quantum mechanical transmission functions at low energies, i.e. around $E_F$, for (c) straight and (d) zigzag DB loops. The dashed lines in both cases indicate the asymmetric connection of the DB loops to the graphene nanoribbons, see also the corresponding insets illustrating the different nanribbon-DB loop-nanoribbon contact configurations.}
  \label{fig_DOS}
\end{center}
\end{figure*}

Upon selectively removing the H atoms from the Si(100) surface to create the DB topologies, localized states emerge in the band gap of Si and the Fermi energy of the system ($E_F$) is shifted towards the band gap edge. It turns out that these localized states include contributions from both surface DB states and up to five layers of the Si substrate underneath (\ref{fig_orb}); as a result, it is important to also include the substrate when computing the charge transport through the system \cite{Joachim_logic}. However, there is no direct tunneling between the loop arms oriented along the dimer rows, unless an electron reaches the connection on top or bottom of the loop.
The electronic DOS of the DB loop depends on the particular topology and it is different for  straight and zigzag loops as shown in \ref{fig_DOS}a and \ref{fig_DOS}b. In the latter case more states, resulting from the symmetry breaking induced by the zig-zag conformation,  appear on both sides of the Fermi energy within an  energy window of roughly $0.2$ eV. This eventually leads to a higher transmission (and current) for this topology, compare panels c and d in \ref{fig_DOS}. In the case of a straight loop the DB states are mainly created below the Fermi energy,  in a smaller energy window  (around $0.16$ eV) closer to the conduction band. The corresponding transmission displays therefore a very asymmetric behavior around the Fermi energy with strong suppression of the conductance already few meV above the Fermi level (\ref{fig_DOS}c). 

Since DB loops form a close contour, one could expect to have a Mach-Zehnder like interferometer supported by the conducting surface states. In a semi-classical picture, electrons that enter the DB loop following different trajectories clockwise or counterclockwise have different phases when they reach the opposite lead. This may result in constructive or destructive interference effects, depending on the path the electron propagates. This effect might be expected only if the DB states take part in the transport, while tunneling between the loop side arms through the substrate will destroy any interference. Such quantum interference effects can be best revealed by varying the contact position between one of the electrodes and the DB loop. On \ref{fig_DOS} we show the effect of symmetric (solid line) and asymmetric (dashed line) coupling of the leads on the transmission function for straight (c) and zigzag (d) DB loops.  In the case of asymmetric coupling,  transport around the Fermi 
energy is dramatically suppressed for both topologies due  to interference effects involving the DB electronic states. The ratio  between the transmissions for the symmetric and asymmetric coupling geometries can be as large as $10^5$. On the contrary, for energies farther away from the Fermi level the transmission is considerably less affected, since the corresponding electronic states have stronger bulk character.  

The interference effects can become more pronounced when  tuning the electron phase by increasing the surface area of the DB loop.
So far we were calculating electronic and transport properties of the full system that includes DB loop  and part of the substrate as  depicted on \ref{fig_syst}. However,  modeling of  larger loops becomes increasingly demanding from the computational point of view, so that instead of inverting large matrices to get the corresponding Green's functions, we use numerically efficient recursive Green's function techniques \cite{metalidis2005green}. Details of the implementation are provided in the SI.
\begin{figure*}[!ht]
\begin{center}
  \includegraphics[width=0.45\textwidth]{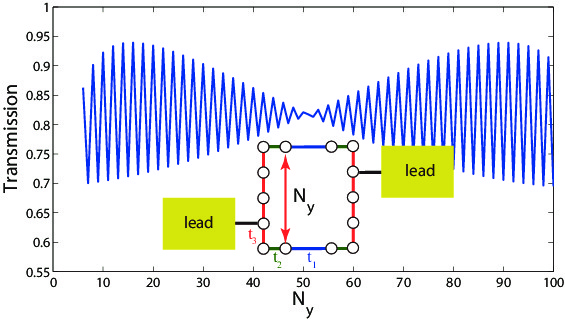}\\
\ \\
\includegraphics[width=0.86\textwidth]{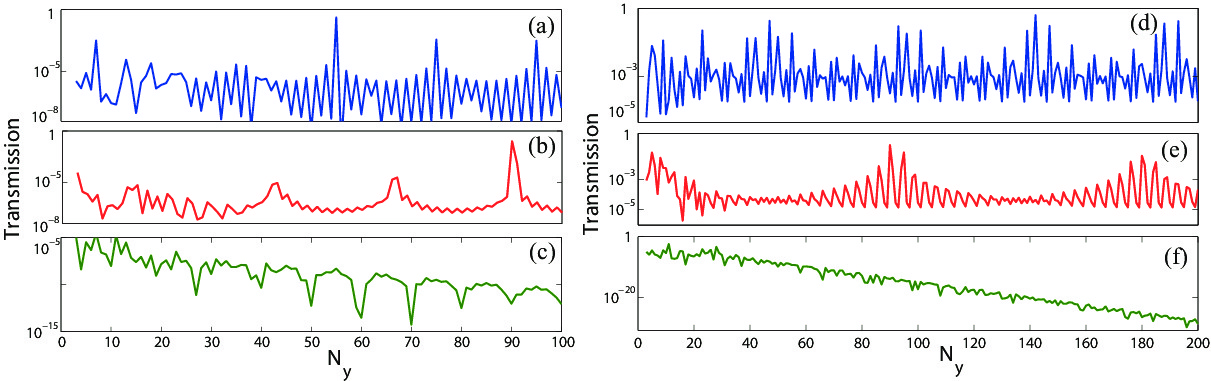}
  \caption{Top panel: Transmission function {at energy $E_F=0.89$} through the toy model loop { with onsite energy equal to 4, measured in terms of hooping parameter between the sites of the semi-infinite squire lattice leads, connected to the loop}, as a function of the loop length $N_y$. Bottom panel: Typical transmission functions through the straight (left) and zigzag (right) DB loops, as a function of the loop length $N_y$, calculated near the Fermi energy at $E-E_F=-0.02 \textrm{ (a) },-0.027 \textrm{ (b) }, -0.01 \textrm{ (c) } -0.05 \textrm{ (d) },0.05 \textrm{ (e) }, 0.0001 \textrm{ (f) }$.}
  \label{fig_toy}
\end{center}
\end{figure*} 

Changing the length of the loop along the (100) direction  is mapped then onto a change in the number of slices of $N_y$ in the recursive Green's function algorithm. Each slice is associated with one DB site and  substrate beneath on each arm of the loop.  Electrons moving clockwise or counterclockwise within the loop gain a different phase  depending on the length. While increasing $N_y$ we move the right lead along the dimer row effectively "reading out" the electron phase difference along the loop. The influence of the loop length can be first illustrated with a simple toy model, see the top panel  of \ref{fig_toy}. The model is described by a nearest-neighbor tight-binding Hamiltonian including only a single state per lattice site, but  different hopping integrals between inequivalent sites: $t_1/t_2=2.1$, $t_2/t_3=1.5$\citep{Joachim_logic}, and connected to  square lattice semi-infinite leads. There are two clear effects  in the transmission curve: even-odd oscillations as a 
function of the loop length, and long-range oscillations that result from constructive (at $N_y=16\textrm{ and }91$) and destructive (at $N_y=51$) interference processes.
%

Going now back to  realistic DB loops, we find for both straight and zigzag topologies the same qualitative behavior as discussed in the toy model, see the bottom panel of \ref{fig_toy}. The odd-even oscillations and long range interference effects are clearly seen if we fix the energy near a transmission resonance, as shown in panels a,b,d,and e in \ref{fig_toy}, while fixing it within a dip region (virtual tunnelling), we see an exponential decrease of the transmission with increasing $N_y$, as illustrated in panels c and f of \ref{fig_toy}. The transmission on panels a,b,d, and e  may differ by more than a factor of $10^5$  depending whether  constructive or destructive interference takes place.  Results concerning  quantum interference  signatures in the $I$-$V$ characteristics of the DB loops with variable size can be found in the Supplementary Material.
%

In summary, we have addressed from an atomistic perspective  charge transport signatures in dangling-bond loops on a Si(100) surface and contacted by mesoscopic graphene nanoribbons. Quantum interference effects have been demonstrated to be  sensitively dependent on the connecting site of the electrodes to the DB loop. Loops built up with a zigzag arrangements of the depassivated atoms along the rows show a by far better low-energy (around the Fermi level) conductance when compared to the straight arrangement. Nevertheless, the strong asymmetry of the transmission around the Fermi level found in the latter case may have interesting consequences for the thermopower of the system, which is, in a first approximation, determined on the behavior of the first derivative of the transmission around the Fermi level of the system. Addressing this issue  requires however a separate study.
Quantum interference effects were also  demonstrated to be only related to DB electronic states, so that their manipulation may allow for an additional control strategy of the electrical transport, since variations in the linear conductance of several orders of magnitude could be realized. This may be implemented by  changing the position of one of the leads, e.g. by using instead of a second graphene nanoribbon a functionalized STM tip, or by applying a voltage between different leads in a multi-terminal setup. Our investigation hints at the possibility of exploiting such effects in loop topologies  for the design of planar-based atomic-scale electronics and its application in e.g. implementing logic 
gates.


This work was partly funded by the EU within the projects
{\textit{Planar Atomic and Molecular Scale devices}} (PAMS, project nr. 610446). This work has also been partly supported by
the German Research Foundation (DFG) within the Cluster of
Excellence "Center for Advancing Electronics Dresden". Computational
resources were provided by the  ZIH at the Dresden University of
Technology. 



Simulation details on  different carbon nanoribbon leads coupled to the DB loops
and technical details related to the recursive Green function approach used to model DBloops with variable size.



\providecommand*{\mcitethebibliography}{\thebibliography}
\csname @ifundefined\endcsname{endmcitethebibliography}
{\let\endmcitethebibliography\endthebibliography}{}


\newpage
\onecolumn 
 \begin{@twocolumnfalse}
\noindent\LARGE{\textbf{Supplementary information}}
\vspace{0.6cm}
\noindent\large{\textbf{Andrii Kleshchonok,$^{\ast}$\textit{$^{a}$} Rafael Guti{e}rrez,\textit{$^{a}$},   and
Gianaurelio Cuniberti\textit{$^{a,b}$}   
}}\vspace{0.5cm}
%
%
\noindent\textit{\small{\textbf{Received Xth XXXXXXXXXX 20XX, Accepted Xth XXXXXXXXX 20XX\newline
First published on the web Xth XXXXXXXXXX 200X}}}
\noindent \textbf{\small{DOI: 10.1039/b000000x}}
\vspace{0.6cm}
%
\noindent \normalsize{ }
\vspace{0.5cm}
\end{@twocolumnfalse}

We provide here additional details on the graphene nanoribbons used as electrodes in the setups discussed in the manuscript. We focus on  the local atomic structure of the electrode-DB contacts and their electronic DOS. We also shortly describe the recursive algorithm used to numerically solve the transport problem for larger loops.  

\begin{figure}
\begin{center}
\includegraphics[width=0.9\textwidth]{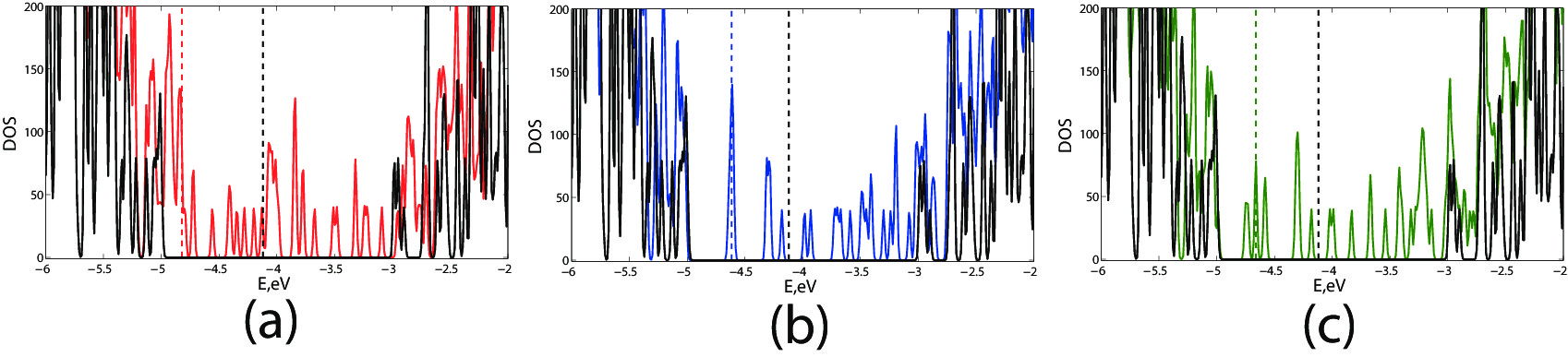}
    \caption{DOS the DB loop coupled to the graphene nanoribbon electrodes that terminate with one (in red, (a), two (in blue, (b)) and three (in green, (c)) C atoms. In black is the DOS of the fully passivated Si. The dotted lines show the corresponding position of the Fermi energy. }
  \label{fig_DOS_trans}
\end{center}
\end{figure}

\begin{figure}[!h]
\begin{center}
\includegraphics[width=0.5\textwidth]{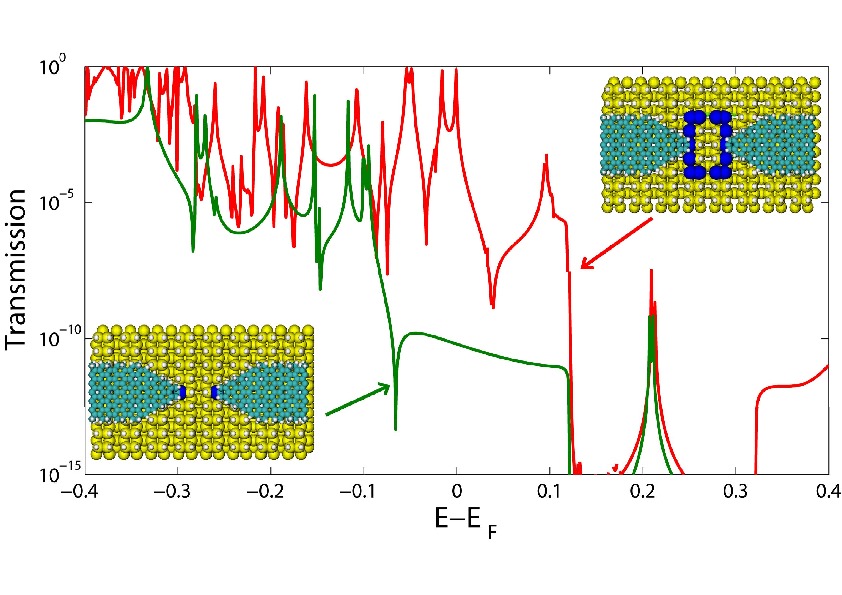}
    \caption{Transmission function around the Fermi energy for the transport within the DB loop (in red) and direct tunnelling between two DBs (in green).}
  \label{fig_2D_trans}
\end{center}
\end{figure}

\begin{figure}
\begin{center}
  \includegraphics[width=0.65\textwidth]{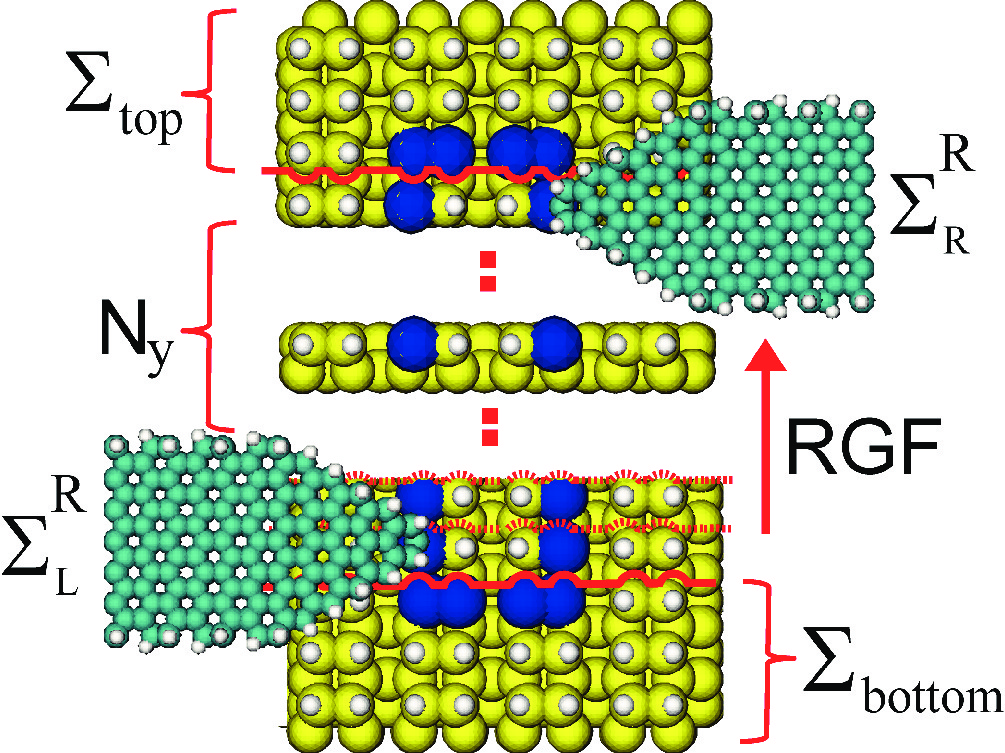}
  \caption{Schematic illustration of the RGF procedure over $N_y$ slices between two leads with self energies $\Sigma^{R}_{L,R}$. Each slice is defined as a part of the system between dashed lines, the solid line shows the part that contributes to the $\Sigma_{\rm{top, bottom}}$.}
  \label{fig_RGF}
\end{center}
\end{figure}

{\textit{Multiple-contact topologies between graphene nanoribbons and DB atoms$-$}} The graphene nanoribbons were put on an H-passivated Si(100) surface at a distance of $3.5 \AA$. The nanoribbon edges are also  passivated with H atoms, with exception of the triangular region, which mediates the electronic contact to the DB loop atoms.  During the relaxation process, which ensures the maximal force component to be below $10^{-7}$ eV/${\AA}$, the electrodes bend towards the closest Si dangling bond atom establishing a connection of the system to the leads with a $1.92 \AA$ Si-C bond distance. This bond  length is in good  agreement with the experimental and simulation data reported before, see for example [Xu, Y.; He, K.; Schmucker, S.; Guo, Z.; Koepke, J.; Wood, J.; Lyding, J.; Aluru, N. \emph{ Nano letters} {\bf 2011}, 11, 2735-2742].\\
The DOS for the different contact geometries  described in the main text is shown in \ref{fig_DOS_trans}. We have considered a straight loop made of  five DBs along the dimer rows and four in the perpendicular direction. In the case of the single-atom contact, case (a),  the position of the Fermi energy is only slightly shifted to $+0.04$ eV with respect to the DB loop alone, while in the cases (b) and (c) the relative change of the Fermi energy amounts to $+0.26$ eV. Moreover, for (b) and (c)  the DOS around the Fermi level contains mainly contributions  from   states localized at the interface between the leads and the Si DBs (b,c). These states are weakly coupled to the rest of the DB loop, having an effective gap of $0.28$ eV. These properties of the electronic structure are also reflected in the transmission function and are responsible for the strong suppression of quantum transport at low energies. These results reveal the way of how the lead should be connected to the DBs. The resulting states should be a mixture of both coming from the leads and DBs or having a small energy gap between them, such that they will be close to the Fermi level.
{In order to demonstrate role of direct electron tunnelling though the substrate we calculate the transmission function of the system made with two DBs on different dimer rows,coupled to the graphene nanoribbon leads. In this case transport of electrons can occur only by direct tunnelling perpendicularly to the dimer rows through the Si substrate.  We compare it with the transmission function through the DB loop \ref{fig_2D_trans}. One can see that around the Fermi energy transmission drops to $10^{-11}$, which is much lower then  in the both cases of destructive and constructive interferences. However around $E-E_F=0.21$ the transmission peak, associated with the direct tunnelling between two DBs, is present in both cases. Exactly only this type of peaks remain in the case of charge localization  on the interface between leads and DB loop discussed above.}
{\textit{Recursive Green's function techniques$-$}} Modeling of larger DB loops is not practically achievable  within a full first-principle framework, so that we use recursive techniques, building  the system slice by slice in the transport direction as it is shown on \ref{fig_RGF}. We construct the Green's function separately for odd and even slices (between the dashed lines in \ref{fig_RGF}) extracting the Hamiltonian and Overlap matrices from the first-principle calculations of the DB loop of smaller sizes. Here we used the assumption, supported  by our DFTB calculations, that the Hamiltonian and Overlap matrices are approximately equal for each odd or even slice. This together with the computed  self-energies $\Sigma^{r}_{\textrm{L,R}}$ and self- energies of bottom and top parts of the system $\Sigma_{\rm{top, bottom}}$ allows us to calculate the transmission function between two leads fully taking into  account the influence of the substrate as well as a  varying number of slices $N_y$ and 
loop geometry (straight or zigzag).

\begin{figure}[ht]
\begin{center}
  \includegraphics[width=0.5\textwidth]{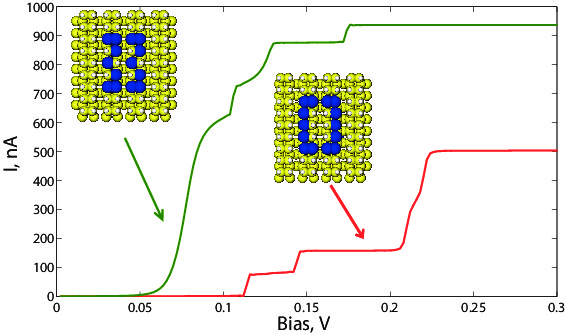}
  \caption{Current as as function of bias voltage through the zigzag (green) and straight (red) DB loops, for a symmetric  coupling of the graphene nanoribbons to the DB loop. }
  \label{fig_cur}
\end{center}
\end{figure}

We finally show in \ref{fig_cur} the $I$-$V$ characteristics of the DB loops with straight and zigzag topologies. The current is computed as   $I(V)=\frac{2e}{h}\int_{-\infty}^{\infty} T(E) \left[ f_{\textrm{R}}(E)-f_{\textrm{L}}(E) \right ]dE$, here $T(E)$ is the zero bias transmission, and $f_{{\textrm{L(R)}}}(E)$ are the left (right) electrode Fermi functions.  As already  predicted for the infinite wires, the current  through the zigzag loop is considerably higher and sets in at lower bias ($\sim 0.05$ V) when compared with the straight loop.   The current mediated by the DB states dominates up to $0.2$ V and $0.15$ V for  zigzag and straight DB loops, respectively; afterwards charge transport through the conduction band starts and the current leaks into the Si substrate.
 
\begin{figure}
\begin{center}
  \includegraphics[width=0.5\textwidth]{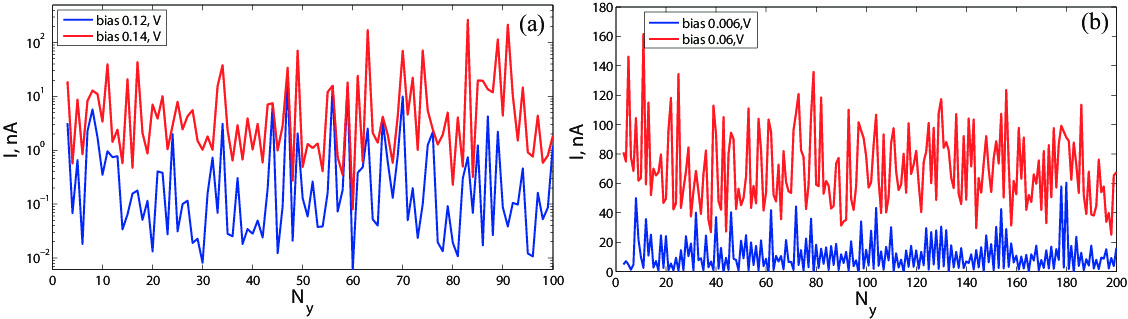}
  \caption{Current  as a function of the DB loop length $N_y$, calculated at bias $0.12$ V and $0.14$ V for the straight (a) and bias $0.006$ V and $0.06$  for zigzag (b) topology.}
  \label{fig_cur_Ny}
\end{center}
\end{figure}

The interference effects are less pronounced in the current vs. loop length plots shown in \ref{fig_cur_Ny}, since by  integrating we average the phases over the bias energy window. However for small voltages, when the DB loops surface states start to conduct, for both loop topologies the quantum interference effects are still noticeable. 

\end{document}